\documentclass[aps,prd,twocolumn,showpacs,superscriptaddress,groupedaddress, nofootinbib]{revtex4} 

\usepackage[utf8]{inputenc}
\usepackage{graphicx}  
\usepackage{dcolumn}   
\usepackage{bm}        
\usepackage{amssymb}   

\usepackage{appendix}

\usepackage{color}

\begin{document}

\title{Improving SDSS cosmological constraints through $\beta$-skeleton weighted correlation functions}

 \author{Fenfen Yin}
 \affiliation{School of Physics and Astronomy, Sun Yat-Sen University, Zhuhai 519082, China}
 \affiliation{School of Data Science, Tongren University, Tongren 554300, China}
 \author{Jiacheng Ding}
 \affiliation{School of Physics and Astronomy, Sun Yat-Sen University, Zhuhai 519082, China}
 \author{Limin Lai}
 \affiliation{School of Physics and Astronomy, Sun Yat-Sen University, Zhuhai 519082, China}
 \affiliation{Department of Astronomy, Shanghai Jiao Tong University, Shanghai 200240, China}
 \author{Wei Zhang}
 \affiliation{School of Physics and Astronomy, Sun Yat-Sen University, Zhuhai 519082, China}
 \author{Liang Xiao}
 \affiliation{School of Physics and Astronomy, Sun Yat-Sen University, Zhuhai 519082, China}
 \author{Zihan Wang}
 \affiliation{School of Physics and Astronomy, Sun Yat-Sen University, Zhuhai 519082, China}
 \author{Jaime Forero-Romero}
 \email{je.forero@uniandes.edu.co}
 \affiliation{Departamento de Física, Universidad de los Andes, Cra. 1 No. 18A-10 Edificio Ip, CP 111711, Bogotá, Colombia}
 \author{Le Zhang}
 \email{zhangle7@mail.sysu.edu.cn}
 \affiliation{School of Physics and Astronomy, Sun Yat-Sen University, Zhuhai 519082, China}
 \affiliation{Peng Cheng Laboratory, No. 2, Xingke 1st Street, Shenzhen 518000, China}
 \affiliation{CSST Science Center for the Guangdong–Hong Kong–Macau Greater Bay Area, SYSU, Zhuhai 519082, China}
 \author{Xiao-Dong Li}
 \email{lixiaod25@mail.sysu.edu.cn}
 \affiliation{School of Physics and Astronomy, Sun Yat-Sen University, Zhuhai 519082, China}
 \affiliation{Peng Cheng Laboratory, No. 2, Xingke 1st Street, Shenzhen 518000, China}
 \affiliation{CSST Science Center for the Guangdong–Hong Kong–Macau Greater Bay Area, SYSU, Zhuhai 519082, China}

\date{\today}

\begin{abstract}
The $\beta$-skeleton approach can be conveniently utilized to construct the cosmic web based on the spatial geometry distribution of galaxies, particularly in sparse samples. This method plays a key role in establishing the three-dimensional structure of the Universe and serves as a tool for quantitatively characterizing the nature of the cosmic web. This study is the first application of $\beta$-skeleton information as weights in mark weighted correlation functions (MCFs), presenting a novel statistical measure. We have applied the $\beta$-skeleton approach to the CMASS NGC galaxy samples from SDSS BOSS DR12 in the redshift interval $0.45 \leq z \leq 0.55$. Additionally, we applied this approach to three COLA cosmological simulations with different settings ($\Omega_m=0.25, \Omega_m=0.31, \Omega_m=0.4$) for comparison. We measured three MCFs, each weighted by i) the number of neighboring galaxies around each galaxy, ii) the average distance of each galaxy from its surrounding neighbors, and iii) the reciprocal of the average distance of each galaxy from its surrounding neighbors. By comparing measurements and calculating corresponding $\chi^2$ statistics, we observe high sensitivity to the cosmological parameter $\Omega_m$ through a joint analysis of the two-point correlation and three MCFs.
\end{abstract}


\maketitle

\section{Introduction}

According to the principles of cosmology, the Universe is considered homogeneous and isotropic at a large scale. However, the combined influence of gravity and the ongoing expansion of the Universe results in a complex and nonlinear structure. In terms of spatial configuration, the unique and intricate web structure formed by the anisotropic collapse of gravity is referred to as the ``cosmic web''~\citep{Bardeen1986Statistics,Lapparent1986Slice,Huchra20122MASS,Tegmark2004three,Guzzo2014VIMOS,Suárez2021Four}.  Topologically, based on the dimensions of gravitational collapse, the cosmic web is typically categorized into knots, filaments, sheets, and voids. 

The study of the structure of the cosmic web relies on large-scale redshift surveys, and currently, there are a number of large-scale observational projects, such as the Sloan Digital Sky Survey (SDSS)~\citep{York2000Sloan,Eisenstein2005Detection,Percival2007Measuring,Anderson2012clustering,Sánchez2012clustering,Sánchez2013clustering,Anderson2014clustering,Ross2015clustering,Beutler2017clustering,Sánchez2017clustering,Alam2017clustering,Chuang2017Linear}, the 2-degree Field Galaxy Redshift Survey (2dFGRS)~\citep{Colless20012dF}, the 6-degree Field Galaxy Redshift Survey (6dFGRS)~\citep{Beutler20126dF,Beutler20116dF}, and the WiggleZ Dark Energy Survey~\citep{Parkinson2012WiggleZ}.

The ongoing stage-IV surveys, such as the Dark Energy Spectroscopic Instrument (DESI)~\cite{2016arXiv161100036D,2022AJ....164..207D}, are expected to further deepen our understanding of the Universe. They possess the capability to observe wider and deeper fields of view with higher resolution, allowing for the collection of more accurate and extensive data and providing additional information about the nonlinear scale structure of the Universe.

The China Space Station Telescope (CSST)~\citep{Gong2019Cosmology}, scheduled for launch in 2025, is a forthcoming stage-IV galaxy survey. It will conduct observations across 17,500 square degrees of the sky, capturing images in the $ugriz$ bands with a spatial resolution comparable to that of the Hubble Space Telescope (HST).

The exploration of the cosmic web has consistently been a significant topic in the study of the Universe. Since early maps of the Universe confirmed the existence of such webs through galaxy redshift measurements, there has been a continuous effort over the past 40 years to find a consistent and stable method to define these weblike structures. Currently, the commonly used analytical tools for studying cosmic webs such as $\beta$-skeleton~\citep{Fang2019skeleton},  FoF \citep{Davis1985evolution}, density-based techniques~\citep{ Klypin1997,Springel_2001,Knollmann2009AhfAH}, T-web~\citep{Hahn_2007, Forero_Romero_2009}, and V-web \citep{Hoffman_2012,Forero_Romero_2014}.

In 2019, the $\beta$-skeleton analysis was employed for the first time to characterize the cosmic web, by analyzing both observational data and accompanying simulations. The observational data utilized the BOSS DR12 CMASS~\citep{SDSS2015Reid} galaxy sample, and the BigMultiDark N-body simulation conducted under the Planck 2015 Cosmology~\citep{Klypin2016MultiDark} served as a test sample. By varying the $\beta$ value and examining the many different scenarios such as redshift evolution and redshift space distortion, it is observed that the number of skeleton connections decreases with increasing $\beta$, resulting in a sparser cosmic web. For instance, when $\beta=1$, numerous connections are generated, irrespective of whether the nodes are located in voids or filaments. However, with setting $\beta=3$, the resulting cosmic web closely resembles that of real observations, indicating a nearly perfect match in the number of connections to the actual galaxies. Finally, at $\beta=10$, only small and relatively isolated compact groups of galaxies are identified and connected. 

Moreover,~\cite{García2020cosmic} explored the information theory entropy of a graph as a scalar to quantify the cosmic web. The findings suggest that entropy can be used as a discrete analogue of scalars used to quantify the connectivity in continuous density fields. The simplicity of computing graph entropy allows its application to both simulations and observational data from large galaxy redshift surveys. This new statistic offers a complementary approach to other topological or clustering measurements. More recently, using data from the IllustrisTNG simulation~\cite{Genel2014Introducing,Vogelsberger2014,nelson2019illustristng,Suárez2021Four} explored a fast machine learning-based approach to infer the underlying dark matter tidal cosmic web environment of a galaxy distribution from its $\beta$-skeleton graph. The $\beta$-skeleton is more suitable for capturing the relatively sparse observed cosmic web compared to traditional T-web and V-web methods. It achieves this by straightforwardly establishing the three-dimensional structure of the Universe based on the spatial geometry of galaxies. In addition, this method overcomes the need for artificially setting a smoothing scale, unlike other methods. It adapts to the density of the environment, generating mathematical topology that traces the details of galaxy distribution in both high- and low-density regions simultaneously. This approach provides a more convenient and feasible way to construct cosmic webs.

The current structure of the Universe has evolved from early small density fluctuations. Through the combined effects of gravitational evolution and cosmic expansion, it gradually formed into the complex structure of cosmic webs observed today, displaying various morphologies at different scales. The traditional and commonly used methods for analyzing the statistical properties of the large-scale structure (LSS) of the Universe include the two-point correlation function (2PCF) or the power spectrum in Fourier space. However, these methods essentially measure the Gaussian content of the density field. With the nonlinear evolution caused by gravitational collapse, the significance of non-Gaussianity increases over time. While higher-order statistics, such as the three-point correlation function~\citep{Sabiu2016Probing,Slepian2017large} and the four-point correlation function~\citep{Sabiu2019Graph}, can capture non-Gaussian information, they face challenges in terms of visual interpretation and efficient calculation.

Recently, the mark weighted correlation function (MCF)~\cite{White:2016yhs}), which assigns density weights to different features of galaxies to extract non-Gaussian information on LSS, has proven effective in capturing more detailed clustering information. MCFs are computationally more feasible compared to other statistical methods capable of extracting non-Gaussian information from LSS. By selecting weights based on the $\alpha$th power of the local density, the weighting amplifies either dense or low-density regions. This method can enhance the distinction between galaxy clusters in dense and sparse regions, resulting in more stringent constraints compared with 2PCF. References \cite{Yang2020Using,Lai:2023dzp} demonstrate that 
 the joint constraint of the MCFs with different $\alpha$ values significantly improved constraints on cosmological parameters such as $\Omega_m$ and $w$. Reference~\cite{Satpathy2019Measurement} presents measurements of MCFs of LOWZ galaxies from SDSS BOSS DR12 to distinguish between the $\Lambda$CDM model and $f(R)$ gravity models. Reference \cite{Xiao2022Cosmological} employed density gradients as weights, and find that the gradient-weighting scheme would produce more robust parameter constraints compared to the density-marked scheme. Furthermore, combining these two schemes yields even stronger constraints than using either one alone.

In this study, we present the first application of the $\beta$-skeleton as  weights in MCFs, serving as a statistical measure of information about the environment contained in the knots of the $\beta$-skeleton. Constructing undirected graph information from the $\beta$-skeleton allows for extracting additional information about LSS. We will apply three weighting schemes to MCFs utilizing undirected graph information derived from the $\beta$-skeleton. These schemes involve weighting by i) the number of neighboring galaxies around each galaxy, denoted as $N_{\rm con}$; ii) the average distance of each galaxy from its surrounding neighbors, denoted as $\bar{D}_{\rm nei}$; and iii) the reciprocal of the average distance of each galaxy from its surrounding neighbors, $1/\bar{D}_{\rm nei}$.

This paper is organized as follows: Sect.~\ref{sect:data} details the COLA and PATCHY simulations, as well as the Sloan Digital Sky Survey III (SDSS-III) Baryon Oscillation Spectroscopic Survey (BOSS) Data Release 12 (DR12) galaxy sample, which are utilized in the subsequent analysis. In Sec.~\ref{sect:method}, we provide a brief introduction to the definition and basic properties of the $\beta$-skeleton, outlining the methodology for calculating the MCFs on the data. 
In Sec.~\ref{sect:result}, we present our findings by applying the MCFs using the weights derived from the $\beta$-skeleton quantities. Finally, we summarize our results and draw conclusions in Sec.~\ref{sect:conclude}.

\section{Data}\label{sect:data}

To investigate the sensitivity of MCF statistics using the properties of the $\beta$-skeleton as the weighting scheme for cosmological parameters, we need to compare actual observations with a series of simulations. Our analysis relies on observations of SDSS BOSS DR12 CMASS galaxies. Additionally, the analysis in this work depends on a set of fast simulations generated using the COmoving Lagrangian Acceleration (COLA) method~\citep{Tassev_2013}. The galaxy mock catalog accurately reproduces the statistical properties of observed CMASS galaxies by employing a recently proposed scheme~\cite{Ding_2024} that utilizes the subhalo abundance-matching (SHAM) procedure and involves adjusting the COLA simulation parameters. Note that to ensure the accuracy of COLA for our analysis, we have compared various statistics with those derived from GADGET simulations using the same cosmological parameters, and found agreement between the two (see the Appendix for details). Additionally, we utilize the Multidark PATCHY simulations (referred to as PATCHY)~\cite{Kitaura2016clustering} to estimate the relevant covariance.

\subsection{Observational data}
The inclusion of the Baryon Oscillation Spectroscopic Survey (BOSS) is a component of SDSS-III~\citep{Eisenstein_2011, Bolton_2012, Dawson_2012}. The detection of the characteristic scale imprinted by baryon acoustic oscillations (BAO) in the early Universe is aimed for by the measurement of the spatial distribution of luminous red galaxies (LRGs) and quasars through BOSS~\citep{Eisenstein_2001}.

Redshift information for approximately $1.5$  million galaxies in a sky region of $\sim 10^{4}$ square degrees is provided by BOSS, which is divided into two samples: LOWZ and CMASS. The LOWZ sample consists of the brightest and reddest LRGs at $z\leq$0.4, while the CMASS sample targets galaxies at higher redshifts, many of which are also LRGs. The sky region covered by CMASS NGC galaxies is approximately $\rm R.A.\in[108^{\circ},\,264^{\circ}]$, $\rm Dec.\in[-4^{\circ},\,69^{\circ}]$. Our study specifically concentrates on a subset of the BOSS DR12 CMASS NGC galaxies within a redshift range of $z\in [0.45, 0.55]$.

\subsection{COLA simulations and mock galaxy catalogs}

COLA,\footnote{https://bitbucket.org/tassev/colacode/src/hg/} a hybrid approach integrating second-order LPT (2LPT) and N-body algorithms, proves to be an effective solution for simulating dark matter (DM) particles. Perturbation theory has demonstrated success in describing large scales, allowing the linear growth rate to replace time integration in N-body simulations. COLA leverages this by employing a comoving frame, with observers following trajectories calculated in perturbation theory. Importantly, COLA accurately calculates large-scale dynamics using 2LPT while assigning the resolution of small scales to the N-body code, without demanding on an exact representation of the internal dynamics of halos. Consequently, COLA can trade accuracy at small scales for computational speed without compromising accuracy at large scales.

To assess the performance of our method and its sensitivity to cosmological parameters, we utilize a series of COLA simulations. To examine the dependence on cosmological parameters, three sets of COLA simulations are employed, each adopting different present-day dark matter density: $\Omega_m=\{0.25,0.31, 0.4\}$, respectively. The simulations are based on a $\Lambda$CDM cosmology with the following parameter values fixed: $\Omega_b=0.048$, $w=-1.0$, $\sigma_8=0.82$, $n_s=0.96$, and $h=0.67$. These values closely approximate the mean constraint derived from the Planck 2015 results~\citep{Ade2016Planck}. We ran COLA simulations with a total of $1024^{3}$ particles in a cubic box, where each side has a length of 800 $h^{-1}$ Mpc.

\begin{figure}
	\includegraphics[width=\columnwidth]{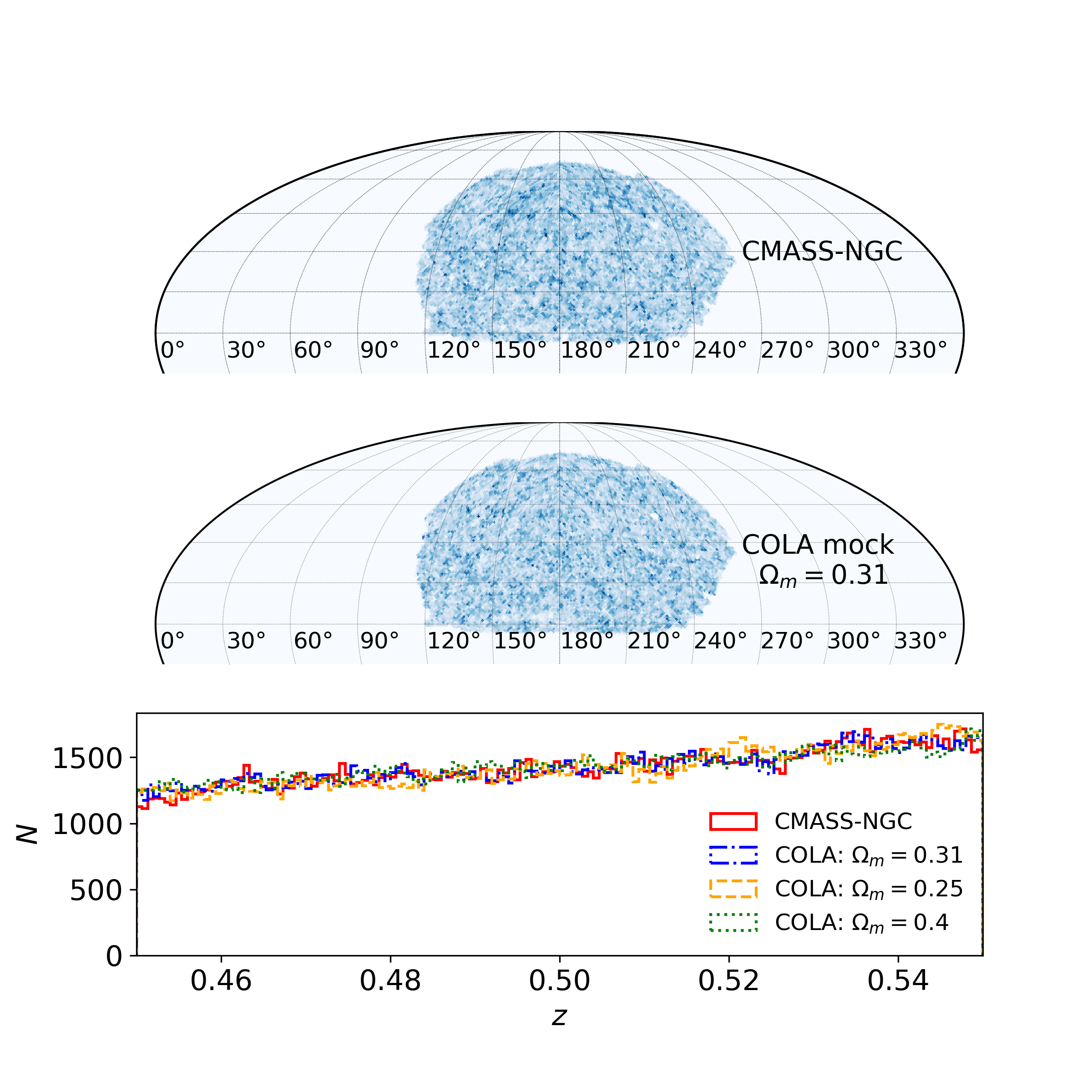}
    \caption{Comparison of the spatial and redshift distributions for observational galaxies and simulation samples. Upper: the R.A.--Dec. distribution of CMASS-NGC galaxies in redshift space is illustrated for the range of $z \in [0.45, 0.55]$. Middle: the R.A.--Dec. distribution of generated COLA mock halos for $\Omega_m=0.31$, showing the same angular distribution as CMASS-NGC galaxies. Bottom: the histograms of the number of galaxies (divided into 150 $z$-bins) are presented for CMASS-NGC (red solid) and COLA simulations with $\Omega_m=0.31$ (blue dashed-dotted), 0.25 (orange dashed), and 0.4 (green dotted), respectively. As seen, the redshift distributions of the three COLA samples closely match that of CMASS-NGC.} 
    \label{fig:radecz}
\end{figure}

For CMASS-NGC galaxies with redshifts in the range of $z \in [0.45, 0.55]$, the spatial and redshift distributions are illustrated in Fig.~\ref{fig:radecz}. The COLA mock survey exhibits a similar angular distribution to that of CMASS-NGC. Additionally, the histograms of the number of galaxies for COLA and CMASS-NGC are displayed at the bottom panel. As shown, the redshift distribution for the simulations and observation are very similar, and the changes in the number of COLA mocks by varying $\Omega_m$ from 0.31 to 0.4 are small.

\subsection{PATCHY simulation for convariance estimation}

To accurately estimate the covariance matrix for MCFs, 1000 PATCHY catalogs~\citep{Kitaura_2016} are employed in this study. The PATCHY (PerturbAtion Theory Catalog generator of Halo and galaxY distributions) mock employs a method that integrates an efficient structure formation model with a local, nonlinear, scale-dependent, and stochastic biasing scheme to produce mock halo catalogs. Augmented Lagrangian Perturbation Theory is utilized for simulating the structure formation, combining 2LPT on large scales with the spherical collapse model on smaller scales. This technique generates a dark matter density field on a mesh starting from Gaussian fluctuations and calculates the peculiar velocity field. In short, PATCHY mock catalogs are generated using approximate gravity solvers and analytical-statistical biasing models, with halo occupation parameters adjusted such that the mocks well reproduce the BOSS two- and three-point statistics. They have been calibrated to the BigMultiDark N-body simulation with high resolution ~\citep{Kitaura2014Modelling,Rodr_guez_Torres_2016}, which utilizes $3840^3$ particles in a volume of $(2.5~h^{-1}\,\rm{Gpc})^3$. This simulation assumes a $\Lambda$CDM cosmology with $\Omega_m = 0.307$, $\Omega_b = 0.048$, $\sigma_8=0.82$, $n_s=0.96$, and $h = 0.67$.

\section{Methodology}\label{sect:method}
\subsection{$\beta$-skeleton analysis}

We will perform the $\beta$-skeleton analysis and use its properties as weights to conduct the MCFs for CMASS-NGC data and COLA mocks. Let us first briefly summarize the $\beta$-skeleton theory and explain how it is used to characterize the cosmic web.

Stemming from the fields of computational geometry and geometric graph theory,  the $\beta$-skeleton algorithm has been widely applied in image analysis, machine learning, visual perception, and pattern recognition~\cite{Edelsbrunner1983On,AMENTA1998125,Zhang2002LocatingSV}. In the realm of web finders, a class of algorithms, including the $\beta$-skeleton, is employed to construct a graph describing the degree of connectedness, starting from a set of 3D spatial points. This process, resembling the minimum spanning tree (MST) algorithm~\cite{Barrow1985Minimal}, has a key distinction – the resulting graph depends on the continuous $\beta$ parameter. Additionally, web finders designed based on topological persistence, such as DisPerSE~\cite{Sousbie_2011}, Betti numbers~\cite{Prana2019Topology} , the caustic skeleton defined based on Lagrangian fluid dynamics \cite{Feldbrugge2017Caustic}, and others, are related to the $\beta$-skeleton. For more detailed information regarding the $\beta$-skeleton in topology and geometric graph theory, please refer to~\cite{Kirkpatrick1985Framework, Correa2013MUTUAL}.

In an $n$-dimensional Euclidean space, the edge set is defined by the $\beta$ skeleton for a point set $S$. In this framework, points $p$ and $q$ in $S$ are considered connected if there is no third point in the various empty regions. Specifically, for $0<\beta<1$, the empty region is formed by the intersection of all spheres with diameter $d_{pq} / \beta$, with $p$ and $q$ on their boundary. When $\beta=1$, the empty region becomes the sphere with diameter $d_{pq}$. For $\beta \geq 1$, the empty region is defined in two different ways: the Circle-based definition and the Lune-based definition, the latter of which is adopted in this study. In this paper, according to this definition, the empty region $R_{pq}$ is the intersection of two spheres with diameter $\beta d_{pq}$, centered at $p+\beta(q-p) / 2$ and $q+\beta(p-q) / 2$, respectively.

The $\beta$-skeleton, as defined above, possesses various interesting mathematical properties. As $\beta$ continuously varies from 0 to $\infty$, the constructed graphs transition from a complete graph to an empty graph. The special case of $\beta=1$ results in the Gabriel graph, known to include the Euclidean MST. In image analysis, for instance, it has been utilized to reconstruct the shape of a two-dimensional object based on a set of sample points on the object's boundary. It has been proven that the Circle-based graphs with $\beta=1.7$ can accurately reconstruct smooth surface boundaries without generating extra edges when samples are dense enough relative to local curvature. It is apparent that the volume of the excluded region exhibits a monotonically increasing relationship with $\beta$. As the excluded region expands, the requirements become more stringent and challenging, making it less likely to find connected paired galaxies. The $\beta$-skeleton code used in this paper is available at the following link.\footnote{https://github.com/xiaodongli1986/LSSCode}

\begin{figure}
	\includegraphics[width=\columnwidth]{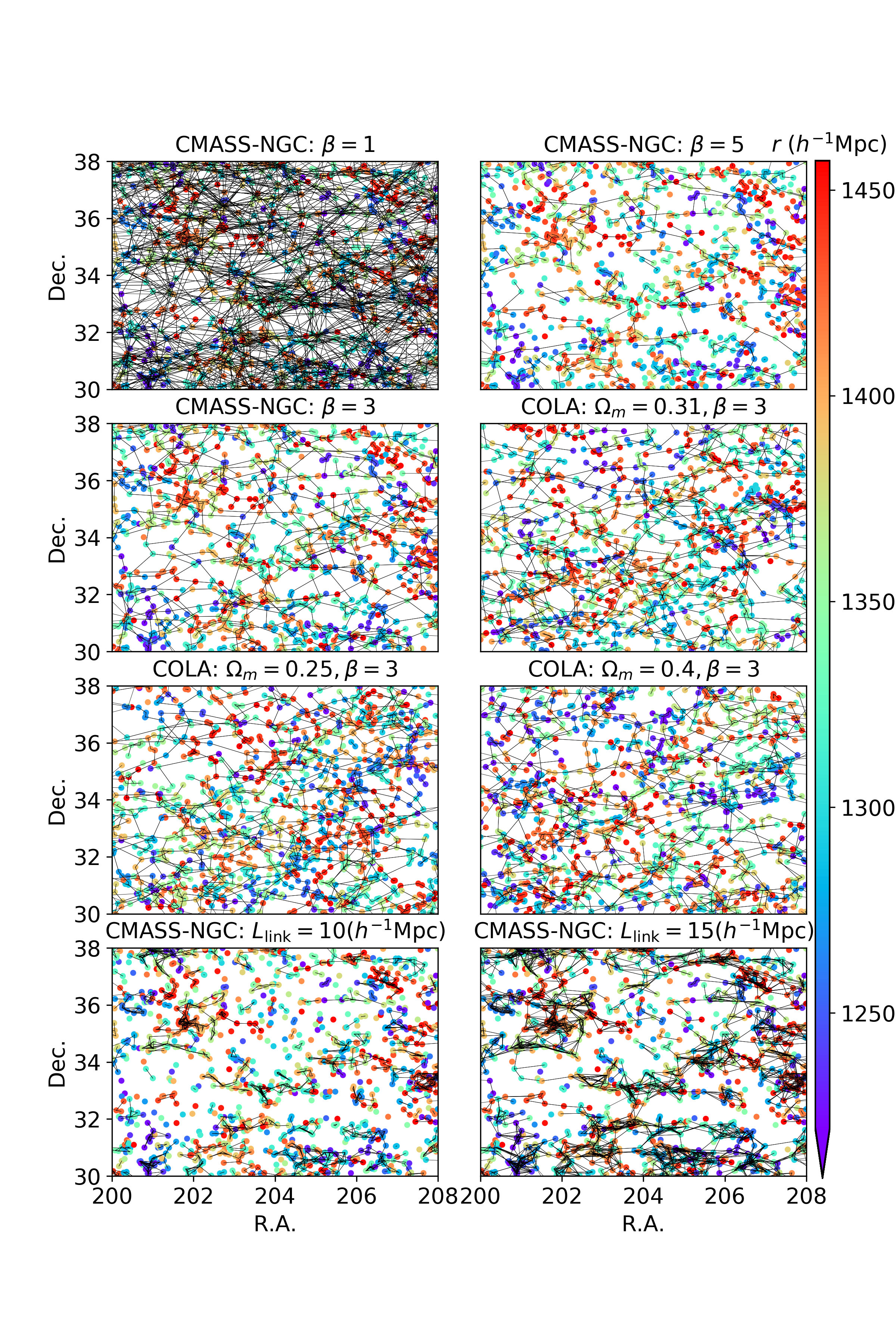}
    \caption{Visualization of cosmic webs produced by the $\beta$-skeleton and Friend-of-Friend (FoF) algorithms for comparison. All maps are selected within the range of R.A. $\in [200^\circ, 208^\circ]$ and Dec. $\in [30^\circ, 38^\circ]$. In the visualization, the scatter represents galaxies, and the resulting connections between galaxy pairs are also depicted. The color is indicative of the comoving distance of the respective galaxies. Upper: the $\beta$-skeleton web of CMASS-NGC galaxies is illustrated for $\beta=1$ (left) and $\beta=5$ (right). The second and third rows: the $\beta$-skeleton web is obtained by choosing $\beta=3$ for CMASS-NGC and COLA simulations with $\Omega_m=0.31$, $\Omega_m=0.25$, $\Omega_m=0.4$. Bottom: the cosmic web obtained from FoF with link lengths, $L_{\rm link}$, set to $10$ (left) and $15$ (right), respectively. In comparison to $L_{\rm link}=$ 10 and 15 in FoF, the $\beta$-skeleton is adaptive to both low and high-density regions. As observed, the $\beta$-skeleton graph is more reliable and natural compared to the cosmic webs produced by FoF, highlighting the advantages of the $\beta$-skeleton graph. $\beta=3$ is found to be optimal and better corresponds to the real cosmic web, represented as the plot of $\Omega_m=0.31,\beta=3$, in contrast to the two graphs that exhibit overly dense ($\beta=1$) or overly sparse ($\beta=5$) patterns.}
    \label{fig:web}
\end{figure}

The $\beta$-skeleton algorithm is applied with $\beta$ values of 1, 3, and 5, respectively, to both CMASS-NGC galaxies and COLA mock samples for various values of $\Omega_m$. The derived cosmic webs are depicted in Fig.~\ref{fig:web}. For comparison, the Friend-of-Friend (FoF) algorithm is applied with linking lengths (denoted as $L_{\rm link}$) set to 10 and 15~$h^{-1}{\rm Mpc}$, and the resulting FoF cosmic webs are also displayed (bottom). All maps are selected within the range of R.A. $\in [200^\circ, 208^\circ]$ and Dec. $\in [30^\circ, 38^\circ]$,  corresponding to a slice of side length  $183.96\times 183.96$ $h^{-1}{\rm Mpc}$,  respectively. The depth of these maps is 227.54 $h^{-1}{\rm Mpc}$, corresponding to the redshift range of $z\in [0.45,0.55]$.

Clearly, the skeletons of maps are observed through connections, and the number of connections decreases when increasing $\beta$, confirming the definition proposed for the $\beta$ skeleton. Specifically, as $\beta$ increases, the empty regions also increase. In essence, the requirement for establishing a connection between two galaxies becomes more stringent. Specifically, considering the CMASS-NGC plot, for $\beta=3$, we detected $1182$ connections, which is very comparable to the number of galaxies ($N_{\rm gal}=1345$) in each CMASS-NGC plot. However, at $\beta=1$, we identified $3610$ connections, significantly exceeding the number of galaxy points. Conversely, for $\beta=5$, we only detected $800$ connections.

Moreover, filamentlike structures are generated from the galaxy sample by the $\beta$-skeleton; this is most clearly detected when $\beta=3$. More densely and sparsely distributed filaments are produced by larger and smaller values of $\beta$, unlike cosmic webs observed in reality. Additionally, ``knots'' within the structure are formed by some of these galaxies, where three or more galaxies are connected. Moreover, there are isolated structures with a relatively small number of group members. Varying $\beta$ strongly influences the overall shape of the skeleton maps. For instance, when $\beta=1$, it roughly corresponds to computing two-point correlations, resulting in numerous connections being generated, irrespective of whether these connections lie within a filament or not. With $\beta=3$, the generated set of structures closely approximates the observed cosmic web, with the number of connections comparable to the actual number of galaxies. In contrast, when $\beta=5$, connections are sparse, as expected, identifying and connecting only the small and relatively isolated compact groups of galaxies. Moreover, statistically, the connection length, denoted as $L_{\rm con}$, varies with $\beta$. As $\beta$ increases, $L_{\rm con}$ tends to decrease and appears more concentrated. This is because the stringent threshold makes connecting two points separated by a large distance more challenging. Also, upon visual inspection, it may be challenging to identify obvious differences in the skeleton graphs corresponding to different values of $\Omega_m$.

From these $\beta$-skeleton cosmic webs, we can quantify several relevant statistical measures. These include the comoving distance ($r$), the connection length of each pair of galaxies ($L_{\rm con}$), the orientation of these connections ($\mu$), and the number of surrounding neighbors of each galaxy ($N_{\rm con}$).

\begin{figure*}
	\includegraphics[width=\textwidth]{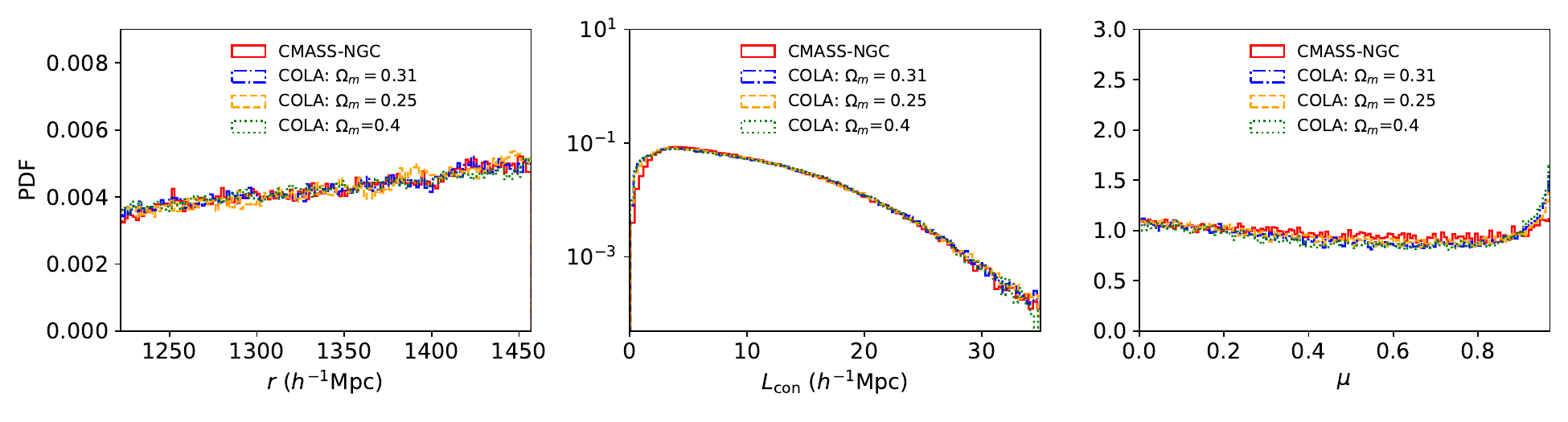}
    \caption{Measured statistics of the $\beta$-skeleton cosmic webs for $\beta = 3$ in both the CMASS-NGC and COLA simulations, considering $\Omega_m$ values of 0.31, 0.25, and 0.4, respectively. Left: the normalized histogram of the comoving distance $r$ of the nodes. Middle: the distribution of the connection length $L_{\rm con}$. Right: the density distribution of the orientation of these connections, defined as $\mu = \cos(\theta)$, where $\theta$ is the angle between LOS and the connection line. As seen, the variation in these statistics concerning $\Omega_m$ is minimal, resulting in the $\beta$-skeleton cosmic web being insensitive solely to this cosmological parameter.}
    \label{fig:rlu}
\end{figure*}

In Fig.~\ref{fig:rlu}, we present the measured statistics of the resulting cosmic webs obtained by applying the $\beta$-skeleton algorithm with $\beta=3$. In the left panel, the distribution of the comoving distance $r$ of the nodes is shown. The red solid line represents the CMASS-NGC data, and for comparison, three different cases in the COLA simulations are considered with $\Omega_m$ values of 0.31 (blue dashed-dotted), 0.25 (orange-dashed), and 0.4 (green-dotted).

The deviations between the real data and COLA simulations, regardless of the $\Omega_m$ values used, are considerably small, with no significant differences observed. The mean comoving distance $r$ is 1346.1 $h^{-1}{\rm Mpc}$ for CMASS-NGC, 1345.7 $h^{-1}{\rm Mpc}$ for $\Omega_m=0.25$, 1346.7 $h^{-1}{\rm Mpc}$ for $\Omega_m=0.31$, and 1344.8 $h^{-1}{\rm Mpc}$ for $\Omega_m=0.4$. These values suggest that when varying $\Omega_m$, the changes in $r$ for COLA are on the order of 0.1\%, making it challenging to rely solely on this quantity for distinguishing between cosmologies.

Similarly, the density distributions of the connection length $L_{\rm con}$ and the orientation of these connections, $\mu$, are illustrated in the middle and right panels. We observe that the discrepancies in either $L_{\rm con}$ or $\mu$ between real data and COLA simulations are statistically insignificant. For instance, the mean $L_{\rm con}$ is 8.83 $~h^{-1} {\rm Mpc}$ for CMASS-NGC, and 8.70, 8.68, and 8.71 $h^{-1}{\rm Mpc}$ for $\Omega_m=0.25, 0.31,$ and $0.4$, respectively, leading to deviations of less than $2\%$ across all cases. Additionally, the mean values of $\mu$, averaged over the range of $\mu\in[0, 0.97]$, are 0.50 for CMASS-NGC, and  0.51, 0.52, 0.53 for $\Omega_m=0.25, 0.31, 0.4$ in COLA, respectively. This suggests that the directions of the connections in all cases are consistently randomly distributed, with no preferred orientation. Moreover, we can clearly observe a rise in density when $\mu>0.9$. This is likely due to the Finger of God (FoG) effect, increasing the orientation along LOS.

\begin{figure}
	\includegraphics[width=\columnwidth]{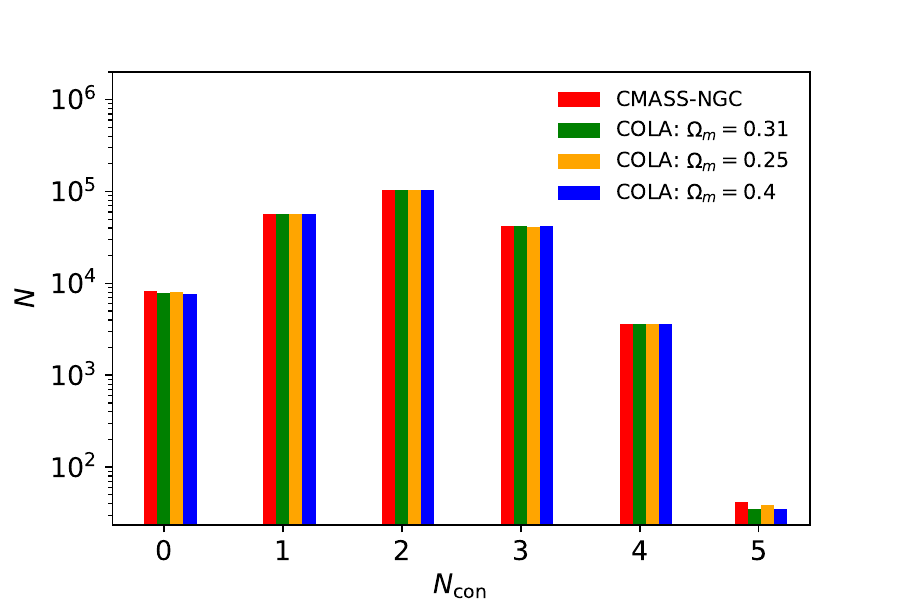}
    \caption{Histogram of the number of neighbors around each galaxy, referred to as connectivity ($N_{\rm con}$), in the range of $[0, 5]$ within the $\beta$-skeleton cosmic webs. It peaks at $N_{\rm con} =2$ and then decreases rapidly. The variation in connectivity is negligible for both CMASS-NGC data and COLA mocks under three different $\Omega_m$ values.}
    \label{fig:connection}
\end{figure}

In Fig.~\ref{fig:connection}, we present a histogram illustrating the connectivity (denoted as $N_{\rm con}$), representing the number of neighbors around each galaxy, displaying the values within the range of $[0, 5]$ for the obtained $\beta$-skeleton cosmic webs.  We observe 8284, 57111, 102639, 41781, 3581  galaxies with $N_{\rm con} = 0, 1, 2, 3, 4$, respectively, along with only 42 galaxies having $N_{\rm con}=5$. In the case of COLA galaxies under $\Omega_m=0.31$, the numbers of galaxies are 7857, 57261, 103680, 41642, 3657, and 35, respectively, for $N_{\rm con}$ ranging from 0 to 5. Notably, the variation in connectivity is found to be negligible across three distinct $\Omega_m$ values for the COLA mocks. Therefore, the sensitivity of $N_{\rm con}$ to $\Omega_m$ is insignificant.

\begin{figure*}
	\includegraphics[width=\textwidth]{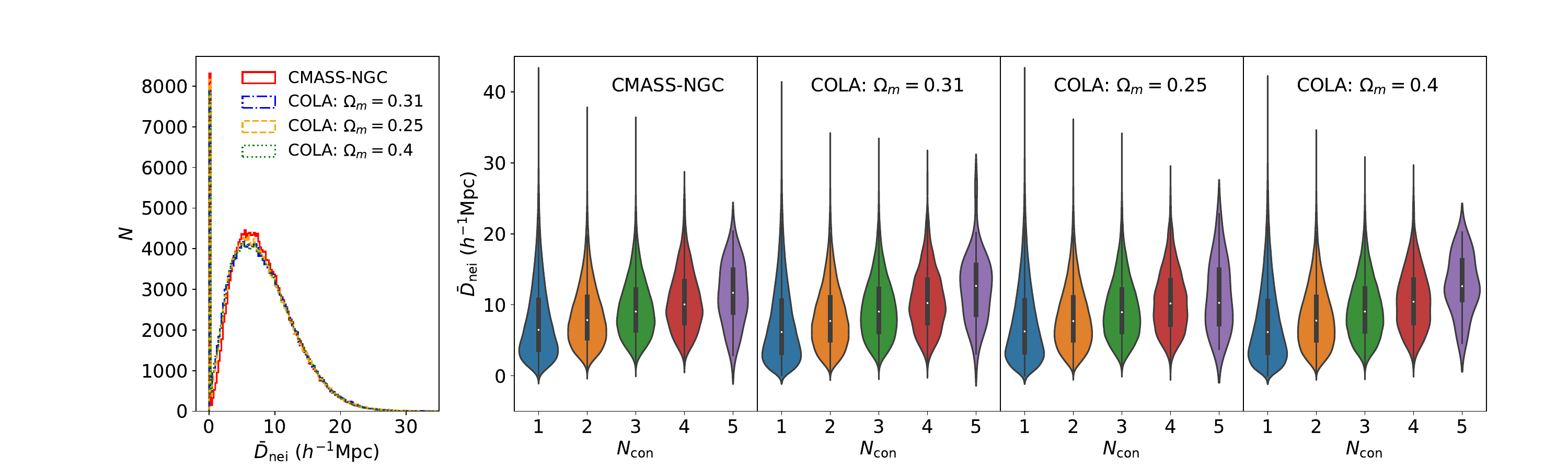}
    \caption{Distribution of the averaged connection length between each galaxy and its surrounding neighbor galaxies, denoted as $\bar{D}_{\rm nei}$, in the case of $\beta=3$, where the neighbor galaxies are determined by their skeleton graph. Left: the histograms display the number of galaxies over $\bar{D}_{\rm nei}$ for both CMASS-NGC data and COLA mocks under three distinct $\Omega_m$ values. The first peak is located at $\bar{D}_{\rm nei}=0$, indicating that galaxies with no connections dominate the cosmic webs for $\beta=3$. The mean value of $\bar{D}_{\rm nei}$ is approximately 8 $h^{-1}{\rm Mpc}$ among all cases. Right (four violin plots): dependence of the distributions of $\bar{D}_{\rm nei}$ on $N_{\rm con}$, ranging from 1 to 5. We present, in sequence, the cases of CMASS-NGC data and COLA mocks for $\Omega_m=0.31, 0.25, 0.4$, respectively.} 
    \label{fig:length}
\end{figure*}

In Fig.~\ref{fig:length}, we illustrate the averaged connection lengths (denoted as $\bar{D}_{\rm nei}$) between each galaxy and its surrounding neighbors, where neighbor galaxies are determined by their skeleton graph in the case of $\beta=3$. On the left side, the histograms (divided into 150 bins) display the number of galaxies over $\bar{D}_{\rm nei}$ for both CMASS-NGC data and COLA mocks across $\Omega_m=0.31$, 0.25, and 0.4, respectively. In all cases, the histograms peak at $\bar{D}_{\rm nei}=0$, indicating that galaxies with no connections dominate the cosmic webs. 

We observe that within the range of $\bar{D}_{\rm nei}$ from 0 to 30 $h^{-1}{\rm Mpc}$ for CMASS-NGC, the total number of galaxies is 213,438. However, beyond this range, the total number decreases exponentially to 86. For the COLA mocks, we find that for $\Omega_m=0.31$, the total number is 214,132 within the range of $\bar{D}_{\rm nei}$ from 0 to 30 $h^{-1}{\rm Mpc}$, and 95 beyond this range. For $\Omega_m=0.25$ and $0.4$, the corresponding totals are 213,428 and 214,472 within the interval of $[0,30]$ $h^{-1}{\rm Mpc}$, and 103 and 69 outside this range. Moreover, the mean value of $\bar{D}_{\rm nei}$ is approximately 8 ${\rm Mpc}/h$ across all cases. 

On the right side, four violin plots depict the dependence of $\bar{D}_{\rm nei}$ distributions on connectivity $N_{\rm con}$, ranging from 1 to 5. The cases of CMASS-NGC data and COLA mocks under three different $\Omega_m$ values are presented sequentially. As seen, all COLA results are nearly identical to the CMASS-NGC case when $N_{\rm con}<5$. However, when $N_{\rm con}=5$, the distribution of $\bar{D}_{\rm nei}$ exhibits slight variation, with mean values  of $\bar{D}_{\rm nei}= 11.52, 12.38, 11.27,12.78~h^{-1}{\rm Mpc}$ for CMASS-NGC and COLA simulations with $\Omega_m=0.31$, $0.25$, and $0.4$.

The above analysis is based on the statistics of the $\beta$-skeleton graphs. We observe that relying on these statistics alone does not enable us to distinguish between cosmologies with significantly different values of $\Omega_m$. Consequently, in the following we will demonstrate that by utilizing the obtained $\beta$-skeleton statistics as weights, the MCF method exhibits high sensitivity to $\Omega_m$.

\subsection{MCFs using the $\beta$-skeleton statistics} 

Next, we will perform an analysis through MCFs using the weights from the $\beta$-skeleton statistics. A standard cosmological analysis generally involves the computation of the 2PCF to infer cosmological information from galaxy clustering properties. Meanwhile, MCFs assign weights to different features of galaxies to extract non-Gaussian information on LSS effectively capturing more detailed clustering information. Considering that the $\beta$-skeleton statistics provide different measures of LSS, we then use these statistics as weights in MCFs to study their sensitivities to the cosmological parameter, the present-day matter density $\Omega_m$. 

In the previous sections, we have introduced several statistical measures for $\beta$-skeleton graph. We will specifically focus on the following three statistics as the weights of MCFs: the number of neighbors around each galaxy, $N_{\rm con}$, the averaged connection length between each galaxy and its surrounding neighbors, $\bar{D}_{\rm nei}$, and its reciprocal, $1/\bar{D}_{\rm nei}$
. Note that, in the following, all MCF analyses will be performed with $\beta$ fixed at 3 for the $\beta$-skeleton measurements. This is the first time to perform an analysis on real and mock LSS data by combining MCFs and the $\beta$-skeleton statistics. 

Briefly, the procedure for calculating the MCFs remains the same as the standard one; however, the weights are assigned based on the measures of the $\beta$-skeleton statistics. Compared with the traditional 2PCF, defined as  $\xi(\boldsymbol{r})=\langle\delta(\boldsymbol{x}) \delta(\boldsymbol{x}+\boldsymbol{r})\rangle$,  the form of MCF, following~\cite{Yang2020Using}, is given by
\begin{eqnarray}
W(\boldsymbol{r})=\left\langle\delta(\boldsymbol{x}) w(\boldsymbol{x})\delta(\boldsymbol{x}+\boldsymbol{r}) w(\boldsymbol{x}+\boldsymbol{r})\right\rangle\,,
\end{eqnarray}
where the term $w(\boldsymbol{x})$ denotes the chosen weights used in MCFs. The term $\delta(\boldsymbol{x})$ denotes the pointlike density contrast, expressed as $\delta(\boldsymbol{x}) = \delta{\rho}/ \bar{\rho}$.

As mentioned earlier, three different weights are chosen for each galaxy in this study, specifically:
\begin{equation}
w(\boldsymbol{x})=\left\{\begin{array}{l}
N_{\rm con}(\boldsymbol{x})\,,\quad\quad {\rm case~I}  \\
\bar{D}_{\rm nei}(\boldsymbol{x})\,,\quad\quad {\rm case~II} \\
1 /\bar{D}_{\rm nei}(\boldsymbol{x})\,,\quad {\rm case~III}
\end{array}\right.
\label{eq:weightrho}
\end{equation}
Here, $w(\boldsymbol{x})$ is obtained from the $\beta$-skeleton at the position $\boldsymbol{x}$. Note that the $\beta$-skeleton graph may have a fraction of samples (detected by knots in the skeleton) compared to the full observed samples of galaxies. Consequently, some galaxies may lack these statistical measurements. In such cases, we assign $w(\boldsymbol{x})=0$. Additionally, when $\bar{D}_{\rm nei}=0$, to prevent divergence for $1/\bar{D}_{\rm nei}$, we also assign $w(\boldsymbol{x})=0$ for case III.

We utilize the widely adopted Landy–Szalay estimator~\cite{Landy1993Bias}, expressed as
\begin{eqnarray}
W(s,\mu) = \frac{WW-2WR+RR}{RR}\,,
\end{eqnarray}
Here, $WW$ denotes the weighted count of galaxy-galaxy pairs, $WR$ corresponds to galaxy-random pairs, and $RR$ represents random-random pairs. These pairs are determined within a distance defined by $s \pm \Delta s$ and $\mu \pm \Delta \mu$, where $s$ represents the distance between pairs, and $\mu$ is defined as $\cos(\theta)$. Here, $\theta$ is the angle between the line of sight (LOS) direction and the line connecting the pair.

The random sample comprises 10 times the number of objects compared to the CMASS-NGC galaxies, and each PATCHY simulation for covariance estimation includes 20 times the number of objects than CMASS-NGC. Based on our testing, results converge with a random sample containing about 10 times more objects than CMASS-NGC, and the weights for the random samples are consistently set to unity.

\subsection{One-dimensional MCFs: $\hat W_0(s)$ and $\hat W_{\Delta s}(\mu)$  }

By integrating $W(s, \mu)$ over either  $s$ or $\mu$, two one-dimensional statistical quantities can be defined. The first one is the monopole of the MCF, a function of the clustering scale, represented as
\begin{eqnarray}\label{eq:ws}
W_0(s) = \int_{0}^{1} W(s, \mu)~d\mu\,.
\end{eqnarray}
The second quantity is the $\mu$-dependent function, represented as
\begin{eqnarray}\label{eq:wmu}
W_{\Delta s}(\mu) = \int_{s_{\min }}^{s_{\max }} W(s, \mu)~ds\,.
\end{eqnarray}
The values $s_{\rm min}=15~h^{-1}{\rm Mpc}$ and $s_{\rm max}=40~h^{-1}{\rm Mpc}$ have been used for quantifying both redshift-space distortions (RSDs) and Alcock-Paczyński (AP) distortions within the context of the tomographic AP method ~\cite{Li_2016}.

To mitigate the impact of galaxy bias and enhance the accuracy of the analysis, it is common to employ normalized statistics, which utilize the shape rather than the amplitude to extract cosmological information. By doing this, the normalized quantities based on Eqs.~(\ref{eq:ws}) and~(\ref{eq:wmu}) can be expressed as follows:
\begin{eqnarray}\label{eq:wsmu}
    \hat{W}_{0}(s) &=& 
    \frac{W_{0}(s)} {\int_a^b W_{0}(s)~ds}\ ,\nonumber \\ 
\hat{W}_{\Delta s}(\mu) &=& \frac{W_{\Delta s}(\mu)}{\int_0^{\mu_{\max}} W_{\Delta s}(\mu)~d \mu} \,.
\end{eqnarray}
In choosing $a = 15~h^{-1}{\rm Mpc}$, $b = 57~h^{-1}{\rm Mpc}$, and $\mu_{\max} = 0.97$ (to reduce effects of fiber collisions and RSDs), extensive testing indicates that the selected parameters are effective for studying clustering features and for enhancing the sensitivity on different cosmologies. From tests, we find that opting for equally spaced divisions of $s$ within the range of $s\in[15, 57]~h^{-1}{\rm Mpc}$ and $\mu$ in the interval $\mu \in [0, 0.97]$ into seven bins results in optimal and robust constraints on $\Omega_m$.

\section{Results}\label{sect:result}

\begin{figure*}
	\includegraphics[width=0.75\textwidth]{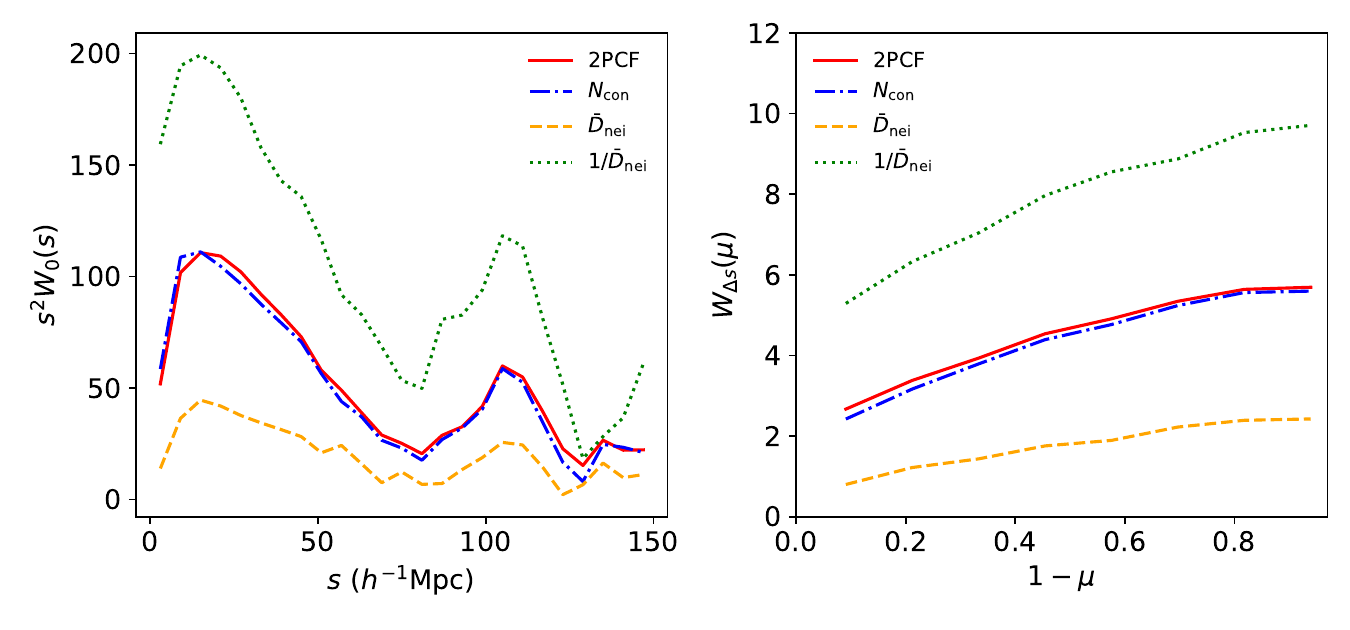}
    \caption{Comparison of one-dimensional MCFs with different weighting schemes for the CMASS-NGC dataset. Left: In comparison to the 2PCF (red solid), the monopole of MCFs, $s^2 W_0(s)$ [as defined in Eq.~(\ref{eq:ws})], is shown for three different weights: $N_{\rm con}$ (blue dashed-dotted), $\bar{D}_{\rm nei}$(orange dashed), and $1/\bar{D}_{\rm nei}$ (green dotted). Right: Same as in the left panel, but for focusing on the anisotropic clustering in MCFs, $W_{\Delta s}(\mu)$ [as defined in Eq.~(\ref{eq:wmu})]. As observed, using the weight $1/\bar{D}_{\rm nei}$ leads to the highest magnitudes in both panels. This is because a number of objects in underdense regions are assigned a larger weight, significantly enhancing the amplitudes of these two-point functions. Conversely, assigning $\bar{D}_{\rm nei}$ as the weight will lead to the lowest amplitudes in both panels. In either the left or right panel, the MCF with $N_{\rm con}$ as the weight results in the amplitude and shape very comparable to those of the 2PCF, implying it is not statistically related to the local properties of each galaxy. }
    \label{fig:mcf}
\end{figure*}

\begin{figure*}
	\includegraphics[width=\textwidth]{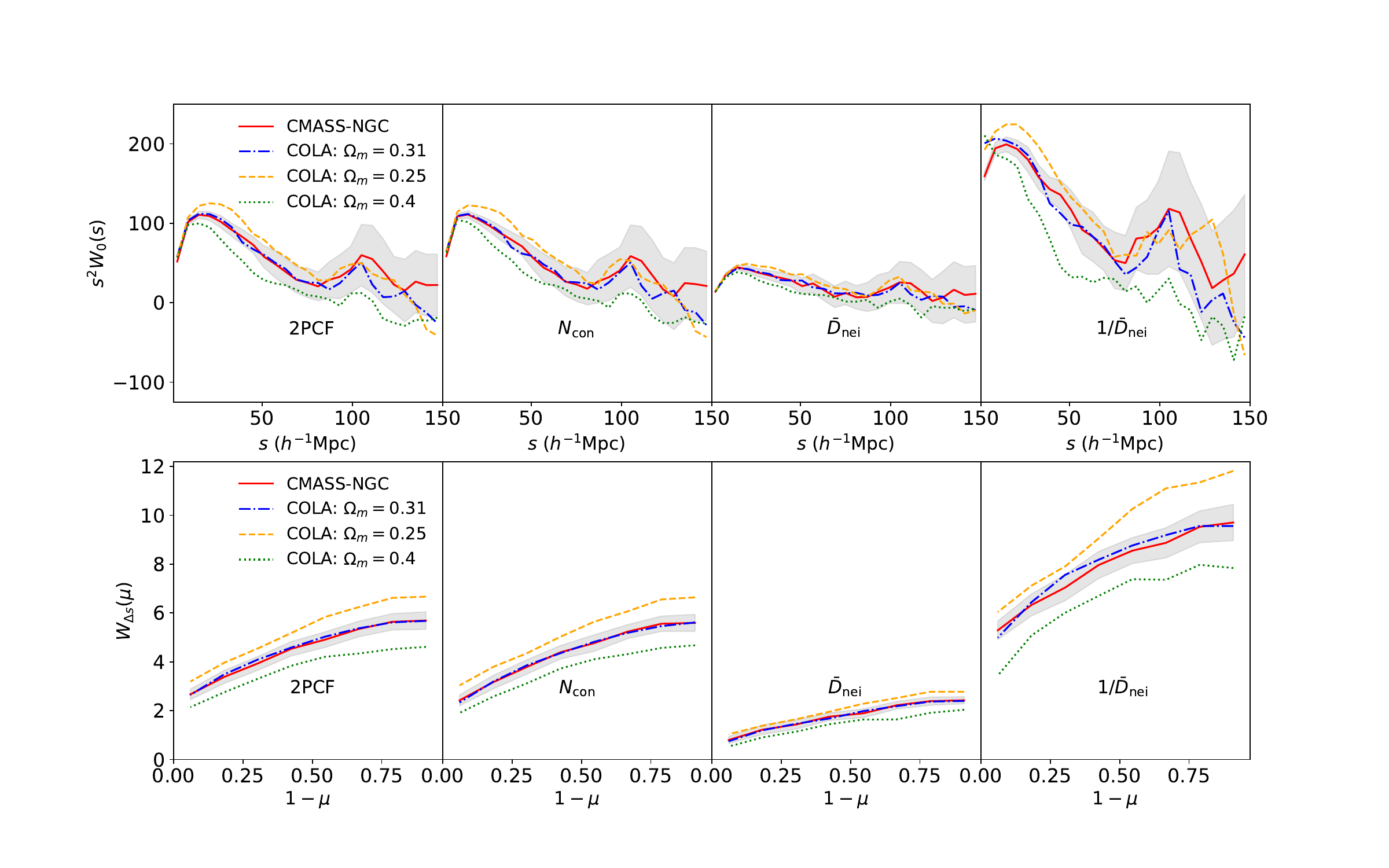}
    \caption{Comparison of one-dimensional MCFs $s^2W_0(s)$ (top) and $W_{\Delta s}(\mu)$ (bottom) for four datasets: CMASS-NGC (red solid), COLA mocks with $\Omega_m=0.31$ (blue dashed-dotted), $\Omega_m=0.25$ (orange dashed), $\Omega_m=0.4$ (green dotted), respectively. From left to right, we present the results for the 2PCF and MCFs using the weights $N_{\rm con}$, $\bar{D}_{\rm nei}$, and $1/\bar{D}_{\rm nei}$, respectively. The gray-shaded region represents $2\sigma$ errors estimated using PATCHY mocks. For the COLA mock with $\Omega_m=0.31$, all derived MCFs in the $s$ and $\mu$ regions we considered ($s \in [0, 150]~h^{-1}\rm Mpc$, $\mu \in [0, 0.97]$) match well with those of CMASS-NGC within $2\sigma$ level. However, when $\Omega_m=0.4$ or 0.25 in COLA largely deviates from the true value in CMASS-NGC, both 2PCFs and MCFs become significantly inconsistent with the observational data beyond $2\sigma$ level.}\label{fig:mcf_cola}
\end{figure*}

\begin{figure*}
	\includegraphics[width=\textwidth]{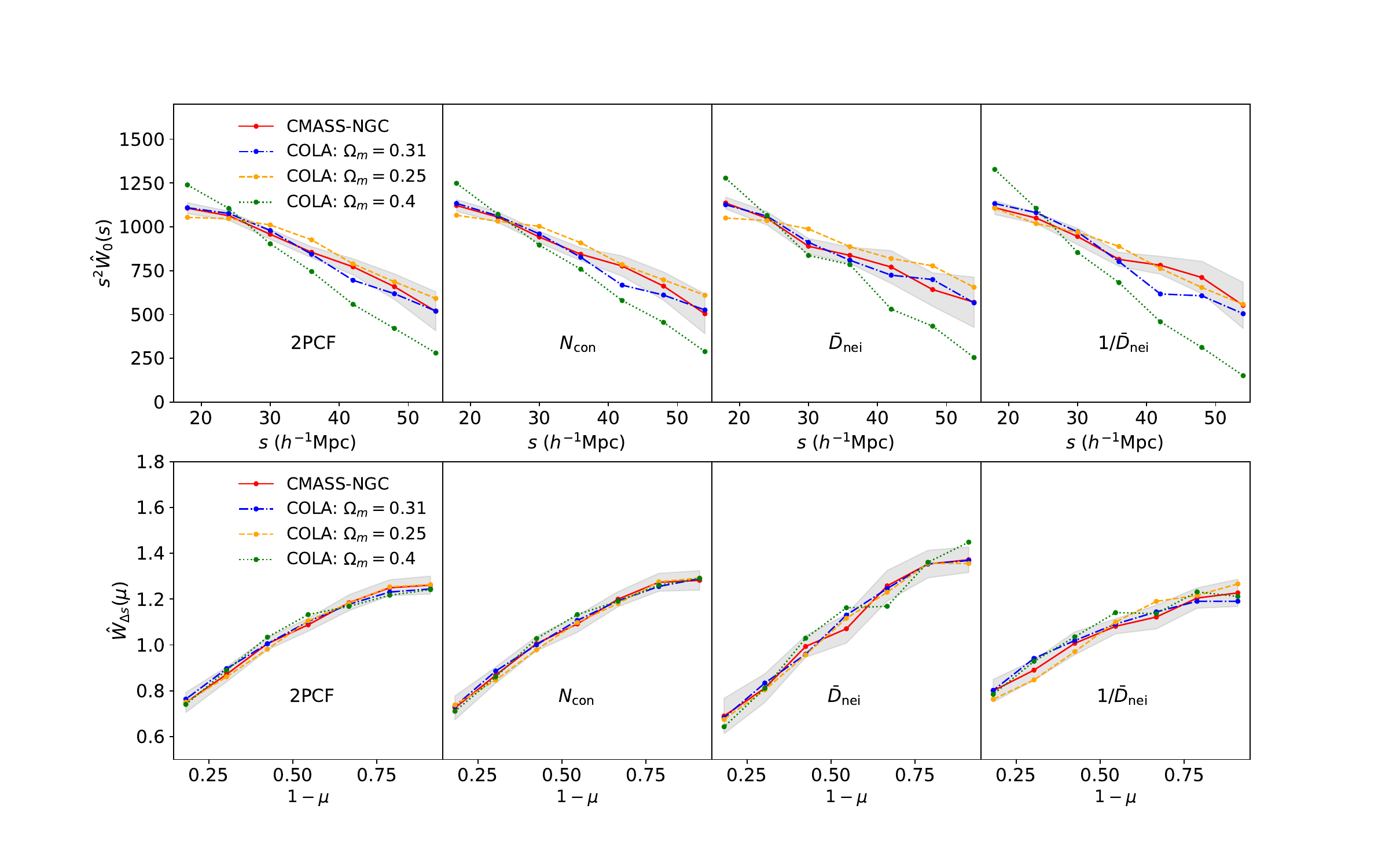}
    \caption{Similar to Fig.~\ref{fig:mcf_cola}, but for the normalized MCFs, $s^2\hat{W}_0(s)$ (top) in the range of $s \in [15, 57]~h^{-1}{\rm Mpc}$, and $\hat{W}_{\Delta s}(\mu)$ (bottom) in the range of $\mu \in [0,0.97]$, equally spaced and divided into 7 bins. The grey-shaded region represents $2\sigma$ errors estimated by the PATCHY mocks. We observe that the results of COLA mocks under $\Omega_m=0.31$ agree well with those of CMASS-NGC, whereas larger and smaller $\Omega_m$ values can lead to apparent deviations, especially for the statistic $s^2\hat{W}_0$. Additionally, we note that the weight $1/\bar{D}_{\rm nei}$ can provide a sensitive probe to $\Omega_m$, as an incorrect value can result in significant deviations from those of the CMASS-NGC data beyond the $2\sigma$ level.}\label{fig:nmcf}
\end{figure*}

Figure~\ref{fig:mcf} illustrates an analysis of the one-dimensional MCFs utilizing various weighting schemes on the CMASS-NGC dataset. On the left panel, contrasting with the 2PCF (red solid), the monopoles of MCFs, $s^2 W_0(s)$ [as defined in Eq.~(\ref{eq:ws})], are illustrated for three distinct weights: $N_{\rm con}$ (blue dashed-dotted), $\bar{D}_{\rm nei}$ (orange dashed), and $1/\bar{D}_{\rm nei}$ (green dotted). The right panel depicts the same comparison but emphasizes the anisotropic clustering in MCFs, $W_{\Delta s}(\mu)$ [as defined in Eq.~(\ref{eq:wmu})]. Notably, the use of the weight $1/\bar{D}_{\rm nei}$ yields the highest magnitudes in both panels. This is attributed to the assignment of a larger weight to numerous objects in underdense regions, significantly enhancing the amplitudes of these MCFs. Conversely, assigning $\bar{D}_{\rm nei}$ as the weight results in the lowest amplitudes. Furthermore, in either panel, employing $N_{\rm con}$ as the weighting scheme for MCFs results in the amplitude and the shape that agree with those of the 2PCF. This is because $N_{\rm con}$ is not statistically related to the clustering strength of the structure (e.g., a galaxy in a low-density region can have a larger $N_{\rm con}$ than a galaxy in a high-density region). Therefore, using it as weights does not significantly alter the statistical clustering.

In Fig.~\ref{fig:mcf_cola}, a comparison of one-dimensional MCFs $s^2W_0(s)$ (top) and $W_{\Delta s}(\mu)$ (bottom) is presented for four datasets: CMASS-NGC (red solid), COLA mocks with $\Omega_m=0.31$ (blue dashed-dotted), $\Omega_m=0.25$ (orange dashed), and $\Omega_m=0.4$ (green dotted), respectively. The results for the 2PCF and MCFs using the weights $N_{\rm con}$, $\bar{D}_{\rm nei}$, and $1/\bar{D}_{\rm nei}$ are displayed from left to right. The gray-shaded region represents $2\sigma$ errors estimated using  PATCHY mocks. For the COLA mock with $\Omega_m=0.31$, all derived MCFs in the $s$ and $\mu$ regions ($s\in[0,150], \mu \in [0,0.97]$) match well with those of CMASS-NGC within the $2\sigma$ level. However, when $\Omega_m=0.4$ or 0.25 in COLA, there is a significant deviation from the true value in CMASS-NGC, causing both 2PCFs and MCFs to become markedly inconsistent with the observational data beyond the $2\sigma$ level.

Moreover, Fig.~\ref{fig:nmcf} illustrates the normalized MCFs, $s^2\hat{W}_0(s)$ (top) within the range $s \in [15, 57]~h^{-1}{\rm Mpc}$, and $\hat{W}_{\Delta s}(\mu)$ (bottom) within the range $\mu \in [0,0.97]$, evenly spaced and divided into seven bins. The gray-shaded region denotes $2\sigma$ errors estimated from PATCHY mocks. The results of COLA mocks under $\Omega_m=0.31$ are well consistent with those of CMASS-NGC. However, deviations become noticeable for larger and smaller $\Omega_m$ values, particularly for the statistic $s^2\hat{W}_0(s)$. Significantly, the weight $1/\bar{D}_{\rm nei}$ serves as a sensitive indicator for $\Omega_m$, where an incorrect value can lead to substantial deviations from those observed in the CMASS-NGC data, exceeding the $2\sigma$ level.

To test the potential improvement of cosmological constraints using $\beta$-skeleton weighting schemes, we select $\hat{W}_0(s)$ and $\hat{W}_{\Delta s}(\mu)$ measurements. We utilize the $\chi^2$ statistic to quantitatively distinguish the wrong cosmologies from the correct one and evaluate the performance in constraining power.
The $\chi^2$ function for fitting the data is given by 
\begin{equation}\label{eq:chi}
    \chi^2=(\Delta\boldsymbol{p})^T\cdot \boldsymbol{C}^{-1}\cdot\Delta\boldsymbol{p}\,,
\end{equation}
where
\begin{equation}
    \Delta\boldsymbol{p}=\boldsymbol{p}_{\rm model}-\boldsymbol{p}_{\rm data}\,.
\end{equation}
Here, $\boldsymbol{p}_{\rm model}=\boldsymbol{p}_{\rm COLA}(\Omega_m)$ represents various statistical quantities of the COLA mocks, including the 2PCF, $\hat{W}_0(s)$, $\hat{W}_{\Delta s}(\mu)$, and their combinations. The symbol  $\boldsymbol{p}_{\rm data}$ denotes the measurements from the observational CMASS-NGC data. By varying $\Omega_m$ within the COLA simulations for $\Omega_m=0.25, \Omega_m=0.31$, and $\Omega_m=0.4$, the resulting $\chi^2$ values are then used to assess the sensitivity of the proposed statistics to $\Omega_m$.

The corresponding covariance matrix, denoted by $\boldsymbol{C}$, is evaluated using the PATCHY simulation mocks. Specifically, the empirical covariance matrix of a vector $\boldsymbol{p}$ is given by
\begin{eqnarray}
\boldsymbol{C} &=& \left<\left(\boldsymbol{p}-\boldsymbol{\bar{p}}\right)\left(\boldsymbol{p}-\boldsymbol{\bar{p}}\right)^T \right> \\
&=&
\frac{1}{N_{\rm mock}-1} \sum_{i=1}^{N_{\rm mock}}  \left(\boldsymbol{p}_i - \boldsymbol{\bar{p}}\right) \left(\boldsymbol{p}_i-\boldsymbol{\bar{p}}\right)^T\,,
\end{eqnarray}
by averaging over all $N_{\rm mock}$ mock samples. Here, $\boldsymbol{p}_i$, with a length of $N_p$, denotes a vector containing all statistical quantities for the $i$th mock sample. Note that the mean of $\boldsymbol{p}$ over all mocks is denoted by $\boldsymbol{\bar{p}}$. We utilized 1000 PATCHY mocks to accurately estimate the relevant covariance matrices for the various considered statistics.

\begin{figure*}
	\includegraphics[width=\textwidth]{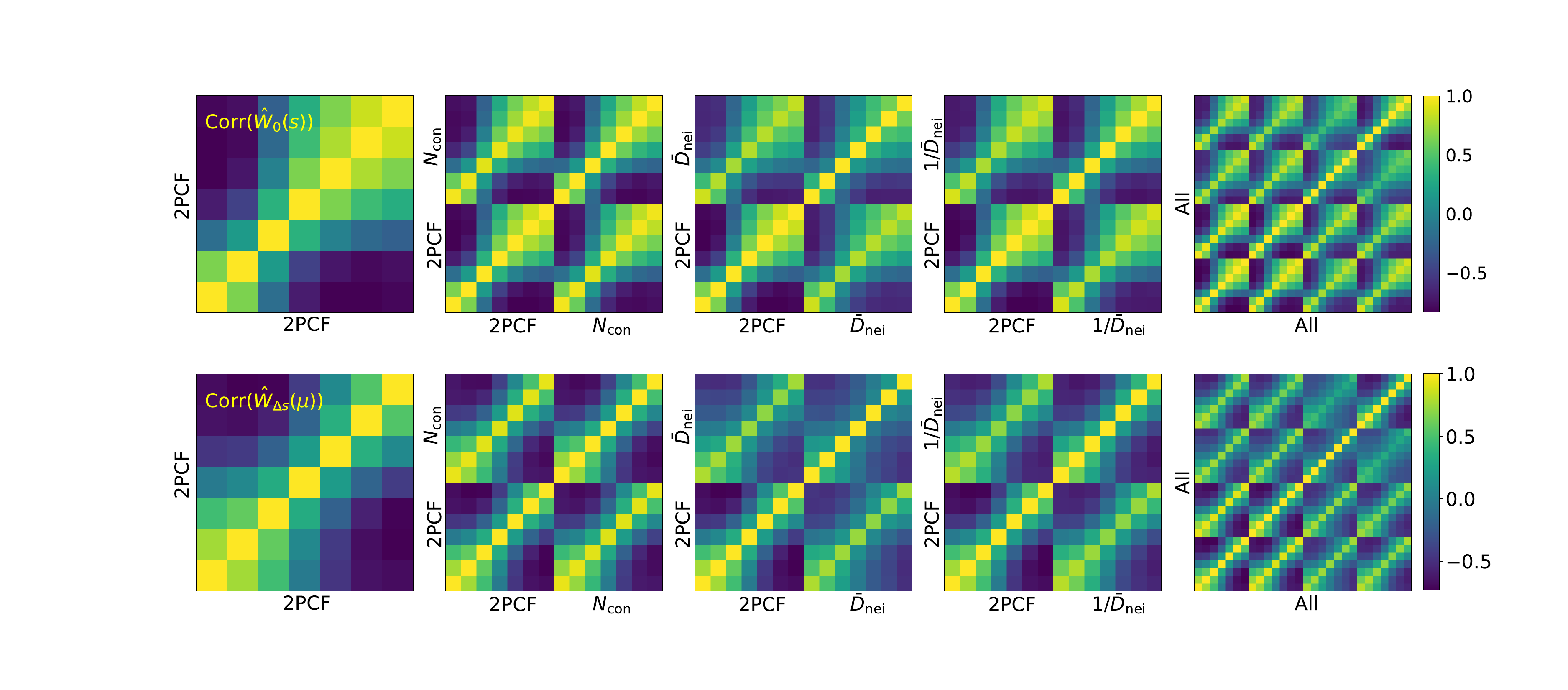}
    \caption{Correlation coefficient matrices of 2PCF,  $\hat{W}_0(s)$ (top)
    and $\hat{W}_{\Delta s}(\mu)$ (bottom). The correlation coefficients are estimated using the 1000 PATCHY mocks, where $\Omega_m=0.31$. To clearly illustrate the correlation strength between different statistics, each matrix displays the cross and auto-correlation coefficients between 2PCF and MCF with three weighting schemes for $N_{\rm con}$, $\bar{D}_{\rm nei}$, and $1/\bar{D}_{\rm nei}$.}
    \label{fig:cov}
\end{figure*}

In Fig.~\ref{fig:cov}, the correlation coefficient matrices of the 2PCF, $\hat{W}_0(s)$ (top) and $\hat{W}_{\Delta s}(\mu)$ (bottom), along with their combinations, are illustrated.  To clearly depict the correlation strength between different statistics, each matrix displays the cross and auto-correlation coefficients between 2PCF and MCFs for three weighting schemes: $N_{\rm con}$, $\bar{D}_{\rm nei}$, and $1/\bar{D}_{\rm nei}$. The correlation between 2PCF and MCFs is notably strong for the weight of $N_{\rm con}$, while such correlations between 2PCF and the other weights become relatively weaker. Additionally, the correlations for combinations of the total of four statistical measurements, including 2PCF and MCFs with the three weights, are shown. These combinations are used to estimate the $\chi^2$ values for the joint analysis.

\begin{figure*}
	\includegraphics[width=0.72\textwidth]{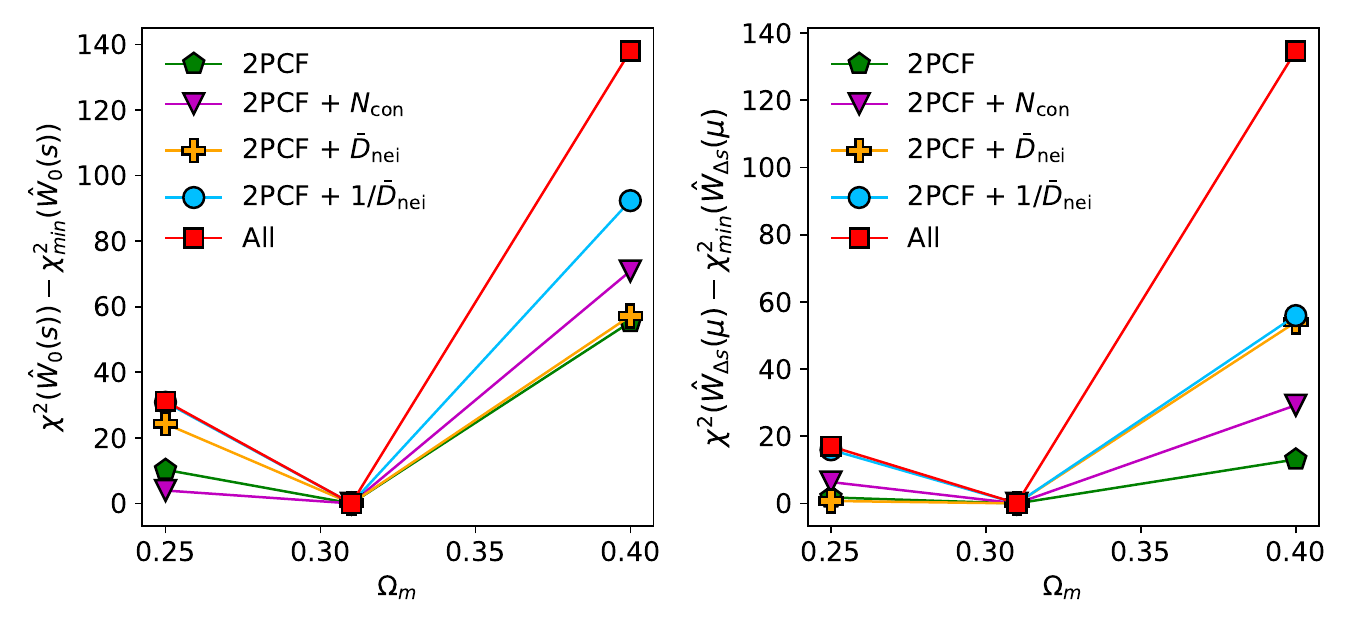}
    \caption{
Changes of $\chi^2$ values [as defined in Eq.~(\ref{eq:chi})] for $\hat{W}_0(s)$ (left) and $\hat{W}_{\Delta s}(\mu)$ (right) as $\Omega_m$ varies from 0.25 to 0.4. Here, $\chi_{\rm min}^2$ is calculated by using $\Omega_m=0.31$ for $\chi^2$, where the $\chi^2$ values reach the minimum for either left and right panel compared to those for the other two $\Omega_m$ values. As seen, compared to the 2PCF, MCFs provide an apparently different dependence of $\chi^2$ on $\Omega_m$. As expected, employing a combination of 2PCF and MCFs with the three $\beta$-skeleton weights results in the highest sensitivity to $\Omega_m$. This is evident from the rapid changes in $\chi^2$ when $\Omega_m$ is varied.}
    \label{fig:chi}
\end{figure*}

To quantitatively assess the sensitivity of various statistical measurements for different $\Omega_m$ values, we further define the change of $\chi^2$ deviating from the same statistical measurement under the fiducial value of 
$\Omega_m=0.31$ (the corresponding $\chi^2$ is denoted by $\chi_{\rm min}$). 
\begin{equation}\label{eq:s/n}
\Delta \chi^2\equiv\chi^2 -\chi^2_{\rm min}\,.
\end{equation}

Displayed in Fig.~\ref{fig:chi} are the changes in $\chi^2$ values, according to Eq.~(\ref{eq:s/n}), for $\hat{W}_0(s)$ (left) and $\hat{W}_{\Delta s}(\mu)$ (right) while $\Omega_m$ ranges from 0.25 to 0.4. The calculation of $\chi_{\rm min}^2$ utilizes the COLA mock with $\Omega_m=0.31$ following Eq.~(\ref{eq:chi}), where the minima in $\chi^2$ values are reached compared to those for the other two $\Omega_m$ values. As observed, MCFs exhibit a notably distinct dependence of $\chi^2$ on $\Omega_m$ compared to the 2PCF. As expected, the use of a combination of 2PCF and MCFs with the $\beta$-skeleton weights yields the highest sensitivity to $\Omega_m$. This is evident from the rapid changes in $\chi^2$ when varying $\Omega_m$.

Moreover, the relative changes of $\Delta \chi^2$ with respect to that of the 2PCF, defined as 
\begin{equation}\label{eq:rs/n}
r= \frac{\Delta \chi^2}{\Delta \chi^2_{\rm 2PCF}}-1\,.
\end{equation}
 The $r$ values are summarized in Table~\ref{tab:chi}. These relative changes can offer another indication of the sensitivity to $\Omega_m$. The measurements involve various combinations of 2PCF and MCFs with three different weights. The comprehensive joint analysis, including all four statistics (labeled as "All (I)" for the combination of 2PCF and three $\hat{W}_0$ statistics, and "All (II)" for 2PCF and three $\hat{W}_{\Delta s}$ statistics), exhibits the highest sensitivity to $\Omega_m$ in either the $\hat{W}_0$ or $\hat{W}_{\Delta s}$ measurements. These relative changes indicate a substantial enhancement in the constraint on $\Omega_m$ through the combination of 2PCF and three MCFs. For instance, the $\chi^2$ values for $\Omega_m=0.25$ are enhanced by 204.06\% for ``All (I)'' and 855.47\% for ``All (II)''. Likewise, when these statistics are combined, enhancements of 149.49\% and 928.33\% are achieved for $\Omega_m=0.40$, respectively.

\begin{table}
\begin{tabular}{lll}
\hline
\hline
Statistics   & $\Omega_m=0.25$ & $\Omega_m=0.40$   \\ \hline
2PCF                           &  0    &  0      \\  
2PCF+$\hat{W}_{0}(N_{\rm con})$        &  -61.61$\%$  & 28.14$\%$    \\
2PCF+$\hat{W}_{0}(\bar{D}_{\rm nei})$  & 136.97$\%$  & 3.33$\%$\\
2PCF+$\hat{W}_{0}(1/\bar{D}_{\rm nei})$ & 200.91$\%$  & 66.94$\%$ \\ 
All (I)             &  204.06$\%$ & 149.49$\%$          \\ 
\hline
2PCF                                  &   0  & 0 \\  
2PCF+$\hat{W}_{\Delta s}(N_{\rm con})$        &    257.52$\%$    & 123.15$\%$ \\ 
2PCF+$\hat{W}_{\Delta s}(\bar{D}_{\rm nei})$   & -58.80$\%$   & 312.19$\%$\\
2PCF+$\hat{W}_{\Delta s}(1/\bar{D}_{\rm nei})$ & 797.33$\%$    & 326.86$\%$\\
All (II)                     & 855.47$\%$    &  928.33$\%$\\
\hline
\hline
\end{tabular}
\caption{Comparison of relative changes of $\chi^2$ with respective to that of 2PCF, i.e. the $r$ value as defined in Eq.~\ref{eq:rs/n}, for different measurements with varying $\Omega_m$ from 0.25 to 0.4. The measurements are based on various combinations of 2PCF and MCFs with three different weights. The overall joint analysis, containing all four statistics (labeled as ``All (I)'' for the combination of 2PCF and three $\hat{W}_0$ statistics and ``All (II)'' for 2PCF and three $\hat{W}_{\Delta s}$ statistics), shows the highest sensitivity to $\Omega_m$ in either the $\hat{W}_0$ or $\hat{W}_{\Delta s}$ measurements. Such relative changes indicate that the combination of the 2PCF and three MCFs leads to a significant improvement in $\Omega_m$ constraint. For instance, the $\chi^2$ values for $\Omega_m=0.25$ are enhanced by 204.06\% for ``All (I)''  and 855.47\% for ``All (II)''. Similarly, combining these two yields enhancements of 149.49\% and 928.33\% , respectively.}\label{tab:chi}
\end{table}

\section{Concluding Remarks}\label{sect:conclude}

The $\beta$-skeleton approach proves to be a convenient tool for constructing the cosmic web based on the spatial geometry distribution of galaxies, especially in sparse samples. It plays a crucial role in establishing the three-dimensional structure of the Universe and serves as a quantitative tool for characterizing the nature of the cosmic web. Moreover, MCFs, by assigning weights to different features of galaxies to extract non-Gaussian information on large-scale structure (LSS), have proven effective in capturing more detailed clustering information.

This study marks the first application of combining mark weighted correlation function (MCFs) and $\beta$-skeleton statistics to real observational data. We have introduced the $\beta$-skeleton information as weights in MCFs, presenting a novel statistical measure. The study applies the $\beta$-skeleton approach to the CMASS NGC galaxy samples from SDSS BOSS DR12 in the redshift interval $0.45 \leq z \leq 0.55$. Additionally, to evaluate the measurements for different cosmologies, we conducted three COLA cosmological simulations with different settings ($\Omega_m=0.25, \Omega_m=0.31, \Omega_m=0.4$) for comparison.

We have measured three MCFs, each weighted by i) the number of neighboring galaxies around each galaxy, $N_{\rm con}$; ii) the average distance of each galaxy from its surrounding neighbors, $\bar{D}_{\rm nei}$; and iii) the reciprocal of the average distance of each galaxy from its surrounding neighbors, $1/\bar{D}_{\rm nei}$. Through a comparison of measurements and the calculation of corresponding $\chi^2$ statistics, we observe a substantial improvement in the constraints on the cosmological parameter $\Omega_m$ by conducting a joint analysis of the standard 2PCF and all three MCFs for different weights. 

In the joint analysis, the highest sensitivity to $\Omega_m$ is observed in either the $\hat{W}_0$ (monopoles of MCFs) or $\hat{W}_{\Delta s}$ (anisotropic clustering of MCFs) measurements. The $\chi^2$ values for the joint analysis are improved by approximately 150\%--928\% compared to the 2PCF alone. Overall, the joint analysis robustly enhances sensitivity to $\Omega_m$, allowing for the rejection of incorrect cosmologies at a significant level. 

Our study has introduced a novel MCF weighting strategy using $\beta$-skeleton information to maximize the extraction of LSS information. This approach holds the potential to be extended to other surveys and datasets, contributing to the constraint of cosmological parameters. In future work, simulation mocks for a wide range of cosmological parameters can be generated through 
emulator approaches, allowing for a thorough assessment of parameter constraints. The method presented here is expected to serve as a valuable analysis tool for upcoming stage-IV surveys,  including the Chinese Space Station Telescope (CSST).

\section*{Acknowledgments}
This work is supported by National SKA Program of China (Grants No. 2020SKA0110401, No. 2020SKA0110402, No. 2020SKA0110100), the National Key R and D Program of China (Grants No. 2020YFC2201600, No. 2018YFA0404504, No. 2018YFA0404601), the National Science Foundation of China (Grants No. 11890691, No. 12203107, 12073088, No. 12373005), the China Manned Space Project with No. CMS-CSST-2021 (( Projects No. A02, No. A03, No. B01), the Guangdong Basic and Applied Basic Research Foundation (2019A1515111098), and the 111 project of the Ministry of Education No. B20019. We also wish to acknowledge the Beijing Super Cloud Center (BSCC) and Beijing Beilong Super Cloud Computing Co., Ltd (\url{http://www.blsc.cn/}) for providing HPC resources that have significantly contributed to the research results presented in this paper.


\section*{APPENDIX}

To ensure that the accuracy of COLA simulation is sufficient for this analysis, here we have made a comparison between the statistics measured in COLA and GADGET~\cite{Springel2001GADGET, Springel2005cosmological, Springel2021Simulating} simulations. GADGET is a massively parallel TreeSPH code for accurate N-body/SPH simulations~\cite{Xu_1995}, which can simulate a collisionless fluid by the N-body method or an ideal gas with smoothed particle hydrodynamics (SPH). \citep{Ding_2024} combines the subhalo abundance matching (SHAM) procedure with COLA and GADGET simulations to create mocks for BOSS CMASS NGC galaxies by fitting the correlation function at $s \in [4, 20]~h^{-1}\rm Mpc$ within the redshift interval $z \in [0.45, 0.55]$. Below we compare the clustering properties of the COLA, GADGET and PATCHY mocks with the observed data.  

To ensure the reliability of our analysis, the COLA and GADGET mocks used the same cosmological parameters $\{\Omega_m=0.2951, \Omega_\Lambda=0.7049, \Omega_b=0.0468, w=-1.0, \sigma_8=0.80,  n_s=0.96\}$ for simulating DM particles and ran simulations under identical conditions with $1024^3$ particles in a box measuring $800~h^{-1}\rm Mpc$.

\begin{figure}
	\includegraphics[scale=0.49]{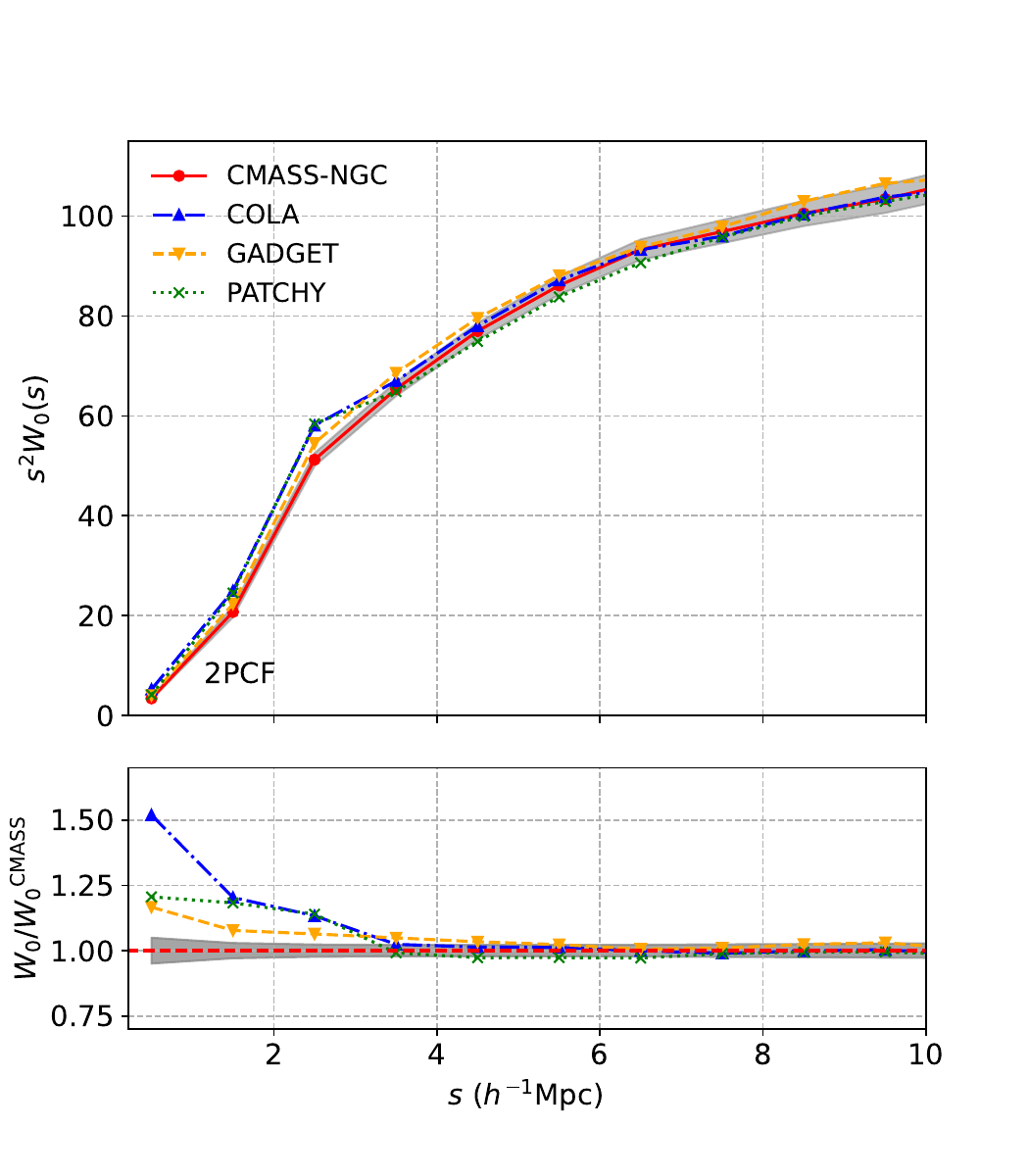}
    \caption{Comparison of 2PCFs $s^2W_0(s)$ for CMASS-NGC (red solid), COLA mock with $\Omega_m=0.2951$ (blue dashed-dotted), GADGET with $\Omega_m=0.2951$ (orange dashed), and PATCHY with $\Omega_m=0.307$ (green dotted), along with their ratios to CMASS-NGC, $W_0/W_0^{\rm CMASS}$ (bottom). The grey shaded region indicates the 2$\sigma$ error estimated from the PATCHY mocks.  2PCFs of COLA and GADGET agree with the data within $2\sigma$ on the fitting scale ($s \in [4, 20]~h^{-1}\rm Mpc$) using the SHAM procedure, while that of PATCHY resides at the boundary of $2\sigma$ due to using the different $\Omega_m$. For scales $s < 4 ~h^{-1}\rm Mpc$, COLA and GADGET deviate slightly more than 2$\sigma$, since this scale exceeds the SHAM fitting range.}\label{fig:mcf_s_10}
\end{figure}

In Fig.~\ref{fig:mcf_s_10}, we present a comparison of the 2PCF $s^2W_0(s)$ (top) obtained from four datasets: CMASS-NGC (red solid), COLA mock with $\Omega_m=0.2951$ (blue dashed-dotted), GADGET with $\Omega_m=0.2951$ (orange dashed), and PATCHY with $\Omega_m=0.307$ (green dotted). We also display their ratios to that of CMASS-NGC, $W_0/W_0^{\rm CMASS}$ (bottom). The gray shaded region represents the 2$\sigma$ error estimated from the PATCHY mocks.

Observably, utilizing the SHAM procedure by fitting the 2PCF in the range of $s \in [4, 20]~h^{-1}\rm Mpc$ to that of CMASS-NGC, both the 2PCFs of COLA and GADGET agree with the data on the fitting scale within the $2\sigma$ level. The COLA and GADGET simulations do not share the same initial condition. Therefore, they have differences caused by the cosmic variance. Due to the use of a different $\Omega_m$ in the PATCHY simulation, 2PCF of PATCHY is essentially on the edge of the $2\sigma$ uncertainty. Furthermore, at smaller scales ($s <4 ~h^{-1}\rm Mpc$), we observe deviations for COLA and GADGET from the data slightly exceeding the 2$\sigma$ level. This discrepancy arises because this scale extends beyond our fitting range in the SHAM procedure.

\begin{figure*}
	\includegraphics[scale=0.54]{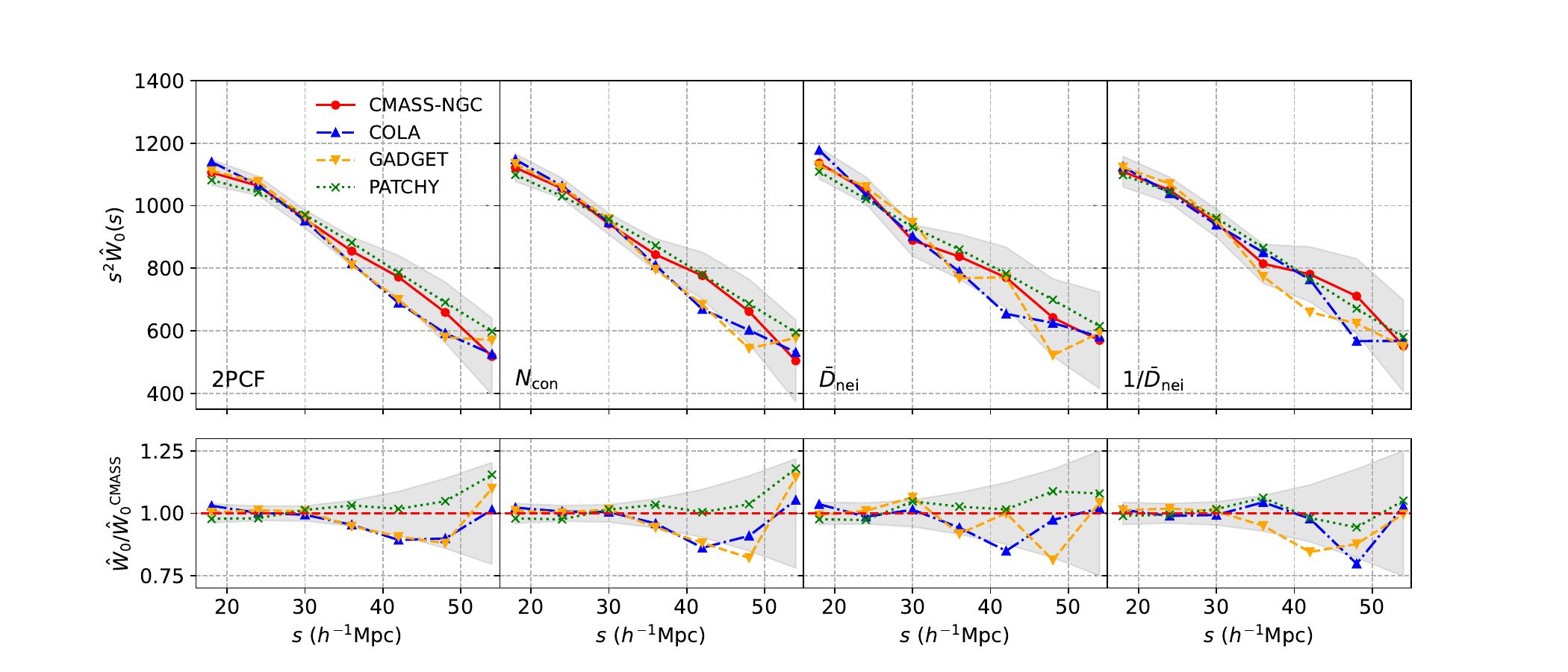}
    \caption{Similar to Fig.~\ref{fig:mcf_s_10}, but for the normalized MCFs, $s^2\hat{W}_0(s)$ (top) and $\hat{W}_0/\hat{W}_0^{\rm CMASS}$ (bottom) in the range of $s \in [15, 57]~h^{-1}{\rm Mpc}$, which are relevant for $\chi^2$ calculations (Eq.~\ref{eq:chi}). The error bars represent $2\sigma$ errors estimated by the PATCHY mocks. COLA and GADGET mocks exhibit good agreement under $\Omega_m=0.2951$, matching CMASS data within about $2\sigma$ level. For all four cases, there is one point slightly exceeding $2\sigma$ in both COLA and GADGET, which can be explained by the cosmic variance. }\label{fig:mcf_s_57}
\end{figure*}

To further validate the consistency of the COLA mock, we conducted a comparison of the normalized 2PCF and MCFs derived from the CMASS-NGC data, GADGET, and PATCHY. This comparison focuses on larger clustering scales within the range of  $s \in [15, 57]~h^{-1}{\rm Mpc}$, which are relevant for the $\chi^2$ calculations [refer to Eq.~(\ref{eq:chi})]. The results are shown in Fig.~\ref{fig:mcf_s_57} to illustrate the normalized MCFs, $s^2\hat{W}_0(s)$ (top) and $\hat{W}_0/\hat{W}_0^{\rm CMASS}$ (bottom). The error bars represent the $2\sigma$ errors, estimated using the PATCHY mocks. 

It is evident that the normalized statistics for the COLA and GADGET mocks exhibit good agreement under the same cosmological parameter $\Omega_m=0.2951$. Both COLA and GADGET simulations closely match the CMASS data within approximately $2\sigma$ level. These statistics from PATCHY are also in agreement with those from the CMASS data at the $2\sigma$ level. Both the COLA and GADGET simulations show one point beyond $2\sigma$ in all four cases, essentially due to the cosmic variance. Thus, we observe a good consistency between COLA and the high-accuracy GADGET simulations, indicating that COLA adequately ensures the reliability of our analysis.

For the stage-III surveys such as SDSS, the accuracy of COLA is sufficient given the associated statistical error of the data sample. However, for the ongoing stage-IV surveys, the statistical error of the sample is greatly reduced. In this case, it is necessary to improve the work of~\cite{Ding_2024} to ensure that the COLA achieves a higher degree of accuracy. Alternatively, one can directly utilize large-scale, high-accuracy cosmological N-body simulations that match the stage-IV surveys such as AbacusSummit~\cite{Maksimova_2021} and Uchuu~\cite{Ishiyama_2021}.

\bibliography{reference}

\end{document}